\documentclass[structabstract]{aa}  
\usepackage{amsmath}
\usepackage{natbib}
\usepackage{graphicx}
\usepackage{txfonts}
\newcommand{\axb}{\mbox{A0620--00}}
\newcommand{\Msun}{\hbox{$\hbox{M}_\odot\;$}}
\newcommand{\Msuno}{\hbox{$\hbox{M}_\odot$}}
\newcommand{\Rsun}{\hbox{$\hbox{R}_\odot\;$}}

\newcommand{\kms}{\hbox{${\rm km}\:{\rm s}^{-1}\;$}}
\newcommand{\kmso}{\hbox{${\rm km}\:{\rm s}^{-1}$}}
\newcommand{\teff}{$T_{\rm eff}\;$}

\newcommand{\logg}{$\log{g}\;$}   	

\bibpunct{(}{)}{;}{a}{}{,} 
\begin{document}
%
%
\title{Doppler tomography of the black hole binary A0620-00 and the origin of
chromospheric emission in quiescent X-ray binaries\thanks{Based on observations obtained with UVES at VLT
Kueyen 8.2 m telescope in programme 66.D-0157(A)}} 
\titlerunning{Doppler tomography and chromospheric emission in A0620-00}
\authorrunning{Gonz\'alez Hern\'andez et al.}
%
%
\author{J.~I.~Gonz\'alez Hern\'andez\inst{1}, \and J.
Casares\inst{2}}
\offprints{J.~I. Gonz\'alez Hern\'andez.}
\institute{
Dpto. de Astrof\'{\i}sica y Ciencias de la Atm\'osfera, Facultad de
Ciencias F\'{\i}sicas, Universidad Complutense de Madrid, E-28040
Madrid, Spain  
\email{jonay@astrax.fis.ucm.es}
\and
Instituto de Astrof{\'\i}sica de Canarias, C/ Via L\'actea s/n, 38200 La
Laguna, Spain
\email{jorge.casares@iac.es}
}

\date{Received ... January 2010; accepted ...}
 
\abstract
{Doppler tomography of emission line profiles in low
mass X-ray binaries allows us to disentangle the different emission sites and
study the structure and variability of accretion disks.} 
{We present UVES high-resolution spectroscopic observations of the
black hole binary \axb\ at quiescence.} 
{These spectroscopic data constrain the orbital 
parameters $P_{\rm orb}=0.32301405(1)$ d and $K_2=437.1\pm2.0$ \kmso.
These values, together with the mass ratio
$q=M_2/M_1=0.062\pm0.010$, imply a minimum mass for the
compact object of $M_1 \sin^3 i = 3.15 \pm 0.10$~\Msuno, consistent
with previous works.} 
{The H$\alpha$ emission from the accretion disk is much
weaker than in previous studies, possibly due to a decrease in disk
activity. Doppler
imaging of the H$\alpha$ line shows for the first time a narrow 
component coming from the secondary star, with an observed
equivalent width of $1.4\pm0.3$~{\AA}, perhaps associated to
chromospheric activity. Subtracting a K-type template star and
correcting for the veiling of the accretion disk yields to an
equivalent width of $2.8\pm0.3$~{\AA}. A bright hot-spot is also 
detected at the position where the gas stream trajectory intercepts
with the accretion disk.}  
{The H$\alpha$ flux associated to the secondary star is 
too large to be powered by X-ray irradiation. It is  
comparable to those observed in RS CVn binaries 
with similar orbital periods and, therefore,
is probably triggered by the rapid stellar rotation. 
}  


\maketitle

\section{Introduction}
\label{introduction}

The black hole binary \axb\ (V616 Mon) is one of the most studied low
mass X-ray binaries (LMXBs), considered as the prototype soft X-ray
transient (SXT). It was discovered in 1975 by the \emph{Ariel 5}
\citep{elv75} satellite during a X-ray outburst where the optical
brightness of the system increased by roughly 6~mag in few days.
One year and several months after the system returned to its
quiescent state at $m_V\sim18.35$~mag. The spectrum of a stellar
counterpart was then identified and classified as a K5V-K7V star
\citep{oke77,mur80}.    
Further spectroscopic observations allowed the determination of the 
orbital period of the secondary star at $\sim7.75$~hr \citep{mar86}
which implied the presence of a black hole of  minimun mass $\sim
3.1$~\Msun \citep{mrw94}.  

The orbital inclination of the system was later estimated from IR
light curves at $\sim41^{\circ}$ implying a primary black hole mass
of $11.0\pm1.9$~\Msun 
(Gelino et al. 2001, see also Shahbaz, Naylor \& Charles 1994).
However, this study adopts a K4V
stellar component with \teff$\sim4600$~K, 300~K cooler than the
effective temperature 
\teff$=4900\pm100$~K derived by \citet{gon04} from
high-resolution optical spectroscopic observations. 
This may affect the required veiling in the IR, and consequently, the
derived inclination and black hole mass \citep{hyn05}. 
Furthermore, \citet{can10} present evidence for substantial disk
contamination in their IR light curves and find $i=51\pm1^{\circ}$ which
translates into a lower black hole mass of $6.6\pm0.3$~\Msun.
On the other hand, contradictory results have been reported using 
low-resolution IR spectra. 
\citet{har07} find a very small or negligible disk veiling at IR
wavelengths whereas \citet{fro07} conclude that it can be 18\% in the H band. 
The latter also argue that \teff$\lesssim 4600$~K is needed to fit their 
observations.
However, we must note here that most studies just adopt a \teff based
on spectral classification, derived through comparison
with low resolution template spectra, without determining
the \emph{true} \teff of the template \citep{hyn05}. A similar
inconsistency has been found for the black hole binary \mbox{XTE
J1118+480} for which \citet{gel06} adopted a \teff$=4250$~K where
\citet{gon06,gon08b} derive a spectroscopic \teff
of~$4700\pm100$~K.  

The accretion disk of \axb\ has been studied in the UV \citep{mhr95}
and the optical \citep{mrw94,oro94}, allowing the investigation of the
inner and outer disk. Both works in the UV and optical seem to agree
that the accretion disk is in a \emph{true} quiescent state. However,
this does not mean that the disk is inactive and probably its
variability appears to be relevant \citep{sha04,can08,can10}. 
\citet{sha04} also suggested that
the accretion disk in \axb\ could be eccentric which may have been
confirmed by more recent observations reported by \citet{nsv08}. As
pointed out by these authors, to determine the definite mass of the
compact object, it is very important to understand the structure and
variability of the accretion disk as exemplified by
\cite{can10}. 

Here we present high-resolution spectroscopy of \axb\ where we 
detect clear emission arising from the secondary star in the
H$\alpha$ Doppler map. This feature has only been observed
before in the systems GU Mus \citep{cas97}, 
Nova Oph 77 \citep{har97}, Cen X-4 \citep{tor02,dav05} and Nova
Scorpii 1994 \citep{sha99}. 
These data also allows us to revisit the orbital parameters of the
system, which we find to be consistent with previous studies. 

\section{Observations}

We obtained 20 spectra of \axb\ with the UV-Visual Echelle
Spectrograph (UVES) at the European Southern Observatory (ESO),
{\itshape Observatorio Cerro Paranal} (Chile), using the 8.2~m {Very 
Large Telescope} (VLT) on 5, 17 and 21 December 2000, covering the
spectral regions $\lambda\lambda$4800--5800 {\AA} and
$\lambda\lambda$5800--6800 {\AA} at resolving power
$\lambda/\delta\lambda\sim43,000$, with a dispersion of 0.029 and
0.035 {\AA}~pixel$^{-1}$ for the blue and red arms, respectively. 
The total  
exposure time was 2.9 hours. The spectra were reduced in a standard
manner using the UVES reduction package within the MIDAS environment.
The exposure time was fixed at $\sim 500$~s to minimize the effects of
orbital smearing which, for the orbital parameters of \axb, is in the
range 2--54 \kmso, and therefore, in some cases, is larger than the
instrumental resolution of $\sim 7$ \kmso. The signal-to-noise
ratio per pixel in the individual spectra is $\sim 4$ and~8 in
continuum regions close to the H$\beta$ and H$\alpha$ lines,
respectively. 
Thus the spectra were binned in wavelength with steps of 0.1
{\AA}~pixel$^{-1}$, increasing the S/N$\sim$6 and~12 at H$\beta$ and
H$\alpha$, respectively.
In this paper, we also use the K3-K4.5V template star HD209100,
observed on 12 November 2000 with the Coralie spectrograph, installed
on the 1.2m Euler Swiss Telescope at the ESO La Silla Observatory
(Chile), with a spectral resolution of
$\lambda/\delta\lambda\sim50,000$. This spectrum was rebinned to the
same step and degraded to the same resolution of the spectra of \axb. 

\subsection{Revised orbital parameters}

We extracted the radial velocities by cross-correlating each UVES
spectrum of the target with the K3-4.5V template spectrum of HD209100,
using the software {\scshape MOLLY} developed by T. R. Marsh. 

Here we have concentrated in the H$\beta$ spectra only, which cover
the spectral regions $\lambda\lambda$4800--5800 {\AA}, because
they contain a larger number of metallic absorption lines for the
cross-correlation and hence provide smaller errorbars by a factor of
2. In any case, the result of cross-correlating the H$\alpha$ spectra
yield identical results. Prior to the cross-correlation, the template
spectrum was rotationally broadened by 96~\kms to match the rotational
velocity of the donor star (see Section~\ref{secrot}).

\begin{figure}[ht!]
\centering
\includegraphics[width=6.5cm,angle=90]{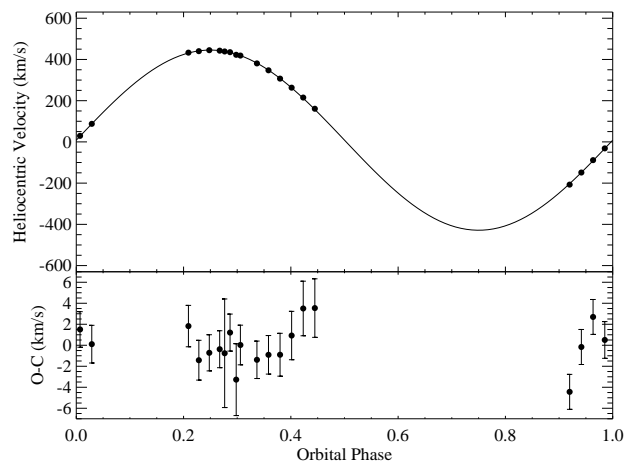}
\caption{\footnotesize{{\it Top panel}: radial velocities of the
secondary star in \axb\ folded on the orbital solution of the data
with best fitting sinusoid. Individual velocity errors are $\lesssim
3$ ${\rm km}\:{\rm s}^{-1}$ and are not plotted because they are
always smaller than the symbol size. {\it Bottom panel}: residuals
of the fit toghether with the individual errors which have been
rescaled by a factor 1.4 so that the minimum reduced $\chi^{2}$ is
1.0.}} 
\label{figrv}
\end{figure}

A $\chi^2$ sine wave fit, $V=K_2\sin [2\pi(t-T_0)/P]$, to the obtained
velocities yields the following orbital solution (see
Fig.~\ref{figrv}):  
$\gamma=8.5\pm1.8$ \kmso, $K_2=437.1\pm2.0$ \kmso, 
 and $T_0=2,451,883.0313\pm0.0002$ d, 
where $T_0$ is defined as the heliocentric time of the 
inferior conjunction of the companion star. The orbital period was 
initially set to the value reported by \citet{mar86} and subsequently 
refined to $P{\rm orb}=0.32301405\pm0.00000001$ d after dividing the 
difference between our $T_0$ and the one quoted by \citet{oro94} by an 
integer number of cycles (17957). 
The quoted uncertainties are at 1$\sigma$ and we have rescaled the 
errors by a factor 1.4 so that the minimum reduced $\chi^{2}$ is 1.0.  

This orbital period, $P$, together with the velocity 
amplitude of the orbital motion of the secondary star, $K_2$, leads to
a mass function of $f(M)=PK_2^3/2\pi G=2.79\pm0.04$ \Msuno. Our value
is consistent at the 1$\sigma$ level with previous results
by \citet[][$f(M)=2.71\pm0.06$]{mrw94} and 
\citet[][$f(M)=2.76\pm0.01$]{nsv08}. 

The derived radial velocity of the center of mass of the
system also agrees at the 1$\sigma$ with previous studies
\citep[e.g. $\gamma=4\pm2$ \kmso,][]{mrw94}. Note that our
high-resolution data has a factor greater than 10 better spectral
resolution than all previous data. For instance, the studies presented
by \citet{mrw94} and \citet{nsv08} used spectra with a resolving power
of 70 and 130 \kmso, respectively. Despite this, \citet{nsv08}
provides a significantly lower uncertainty for the semiamplitude
velocity of 0.5~\kms given their resolving power. 
This is because our orbital coverage is much less complete than
previous studies, specially around orbital phase 0.75.

\subsection{Rotational velocity\label{secrot}}

Following \citet{mrw94}, we also computed the optimal $v \sin i$ by
subtracting broadened versions of the template star, in
steps of 1 \kmso, and minimizing the residual. We used a spherical
rotational profile with linearized limb-darkening $\epsilon = 0.81$, 
which is appropriate for the stellar parameters of the donor star 
\citep{gon04} and our wavelength range $\sim 5250$~{\AA} 
\citep{al78}.  
The best fit corresponds to a $v \sin i=96\pm0.8$
\kmso but the error is purely formal since it does not account for 
systematics due to our choice of $\epsilon$ which is adecuate for the
continuum but likely overestimated for the  
absorption lines \citep{col95}. In an attempt to derive a more
realistic error, we have also computed the rotational broadening for
the extreme case $\epsilon=0$ and find $v \sin i = 87 \pm 0.7$
\kmso. A more conservative value is then provided by the mean of the
two determinations i.e. $v \sin i = 92 \pm 5$ \kmso. As a result of
the optimal subtraction we also find that the donor star contributes
$\sim 80$\% to the total light in the H$\beta$ region and $\sim$85\%
in the H$\alpha$ region. At this point we note that the absorption
features of the secondary star can be smeared by as much as 53 \kmso
according to the length of the exposure times and the orbital phase of
the observations. We have also tried to correct for this by simulating
the smearing in the template spectrum according to \citet{cas97} but
found that the effect is completely negligible.  

In the case of tidally locked Roche lobe filling stars, the
rotational velocity relates to the velocity semiamplitude, $K_2$, and
the mass ratio, $q=M_2/M_1=K_1/K_2$, through the expression 
$v \sin i \simeq 0.462 K_2 q^{1/3} (1+q)^{2/3}$ \citep[e.g.][]{wah88}. 
Our derived rotational velocity, combined with our value of $K_2$,
implies a binary mass ratio $q=M_2/M_1=0.081\pm0.010$. 
However, due to the Roche lobe geometry, $v \sin i$ displays a
$\sim$10\% modulation with maxima at the orbital quadratures
\citep{cas96}.  Most of our spectra are located at orbital 
phases around 0.25 and hence our previous determination is likely to
be overestimated by a factor $\sim$5\% which would make it consistent
with \citet{mrw94} and \citet{nsv08}. Therefore,
we think that the value provided by \citet{mrw94}, $v \sin
i=83\pm5$, is more realistic. Using this result and our value for
$K_2$, we obtain $q=0.062\pm0.010$, which is the same value given in
\citet{nsv08}. This, combined with the mass function, 
provides a minimum mass for the compact object of $M_1 \sin^3 i =
f(M)*(1+q)^2= 3.15 \pm 0.10$~\Msuno.

\begin{figure}[ht!]
\centering
\includegraphics[width=6.5cm,angle=90]{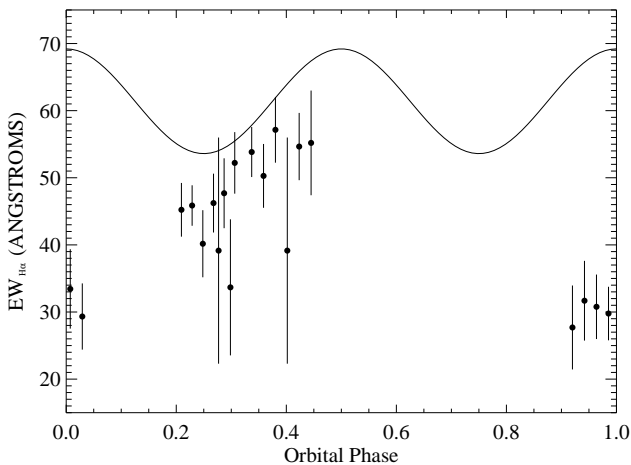}
\caption{\footnotesize{Equivalent width of H$\alpha$ profiles of the
20 UVES spectra of \axb\ versus orbital phase of the secondary star.
The solid line is the solution of \citet{mrw94} for comparison.}}    
\label{figew}
\end{figure}

\section{Ellipsoidal variations of the H$\alpha$ line\label{secew}}

We have determined the equivalent width (EW) of H$\alpha$ profiles for
the 20 UVES spectra of \axb\ by adding numerically the normalized
fluxes multiplied by the wavelength step. 
In Fig.~\ref{figew} we depict the
evolution of the EW of H$\alpha$ with orbital phase. 
The error bars were determined by measuring the change in EW 
from varying the continuum location according to the S/N ratio
in the continuum regions.
Despite our limited phase coverage, the orbital variation of the EWs 
hints at the ellipsoidal varibility seen by \citet{mrw94} and
\citet{nsv08} due to the dominance of the donor star's light.
 
We have also displayed for comparison the solution of the fit to a
sinusoid plus a constant to the data of \citet{mrw94}, 
${\rm EW} ({\AA})= 61.4 + 7.8 \cos 4\pi\phi$. 
Our mean EW is 43 \AA, significantly lower than 
in \citet{mrw94}, whose spectroscopic data were obtained in December
1991, and much lower than that in \citet{nsv08}, whose data were
obtained in December 2006. This may be due to an increase in the 
continuum from the accretion disk, a decrease in the H$\alpha$ flux or
a combination of the two.  
Despite being in quiescence, \axb~ shows significant
variability and different states associated with changes in disk
activity \citep[e.g.][]{can08}. Therefore, the lower EW seen in our
spectra with respect to previous studies could well be associated to a
different level of accretion disk activity. 
The relative contribution of the accretion disk to the
continuum in the H$\alpha$ region is $\sim$15\% in our data and
$\sim$6\%  in \citet{mrw94}. This, combined with the variation in EWs
between the two data sets, implies that the H$\alpha$ flux has dropped
$\sim$ 33\% in our data with respect to \citet{mrw94}. In fact,
\citet{can08} find that \axb~ was in a lower level of activity (what
they called ``passive'' state) most of the time between December 1999
and December 2003. Our observations were taken in December 2000,
when the system was in the ``passive'' state and this seems to be the reason for 
the low level of H$\alpha$ flux that we see.

\section{Doppler images of H$\alpha$ and H$\beta$}

\begin{figure*}[ht!]
\centering
\includegraphics[width=15.5cm,angle=-90]{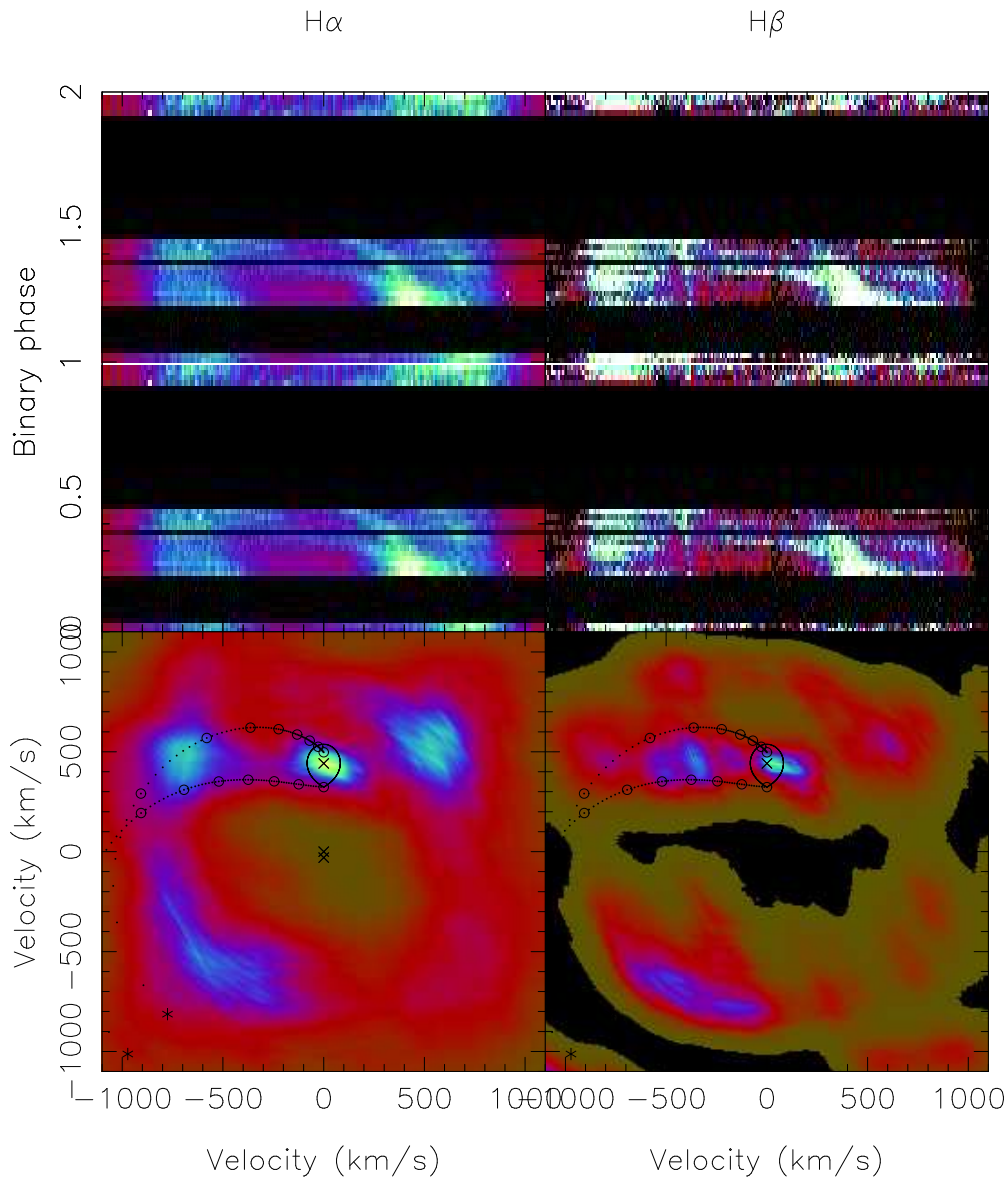}
\caption{\footnotesize{Doppler images of H$\alpha$ and H$\beta$ of the
UVES spectroscopic data. The trailed spectra are displayed in the top
panels where several gaps in phase appear because the observations
were taken in three different nights. Doppler images are depicted
in the bottom panels. The predicted velocities of the gas stream
(which starts from the inner Lagrangian point) and of the disk along
the gas stream are plotted for $K_2=437.1$~\kms and
$q=M_2/M_1=0.062$. Asterisks mark turning points in distance
from the compact object. Circles have been plotted every
0.1~$R_{L1}$, and dots every 0.01~$R_{L1}$ along the streams.}}   
\label{figmap}
\end{figure*}

We have used Doppler tomography \citep{mah88} to map the H$\alpha$ and
H$\beta$ emission in \axb. By combining the orbitally-resolved line
profiles we are able to reconstruct the brightness distribution of
the system in velocity space. The result is displayed in
Fig.~\ref{figmap}. The Doppler maps were built by combining the 20
profiles. The spectra of H$\alpha$ and H$\beta$ were continuum
subtracted, leaving the pure line emission, and rebinned to a velocity
scale of 9 \kmso per pixel. The top panels show trailed spectra and
the bottom panels the corresponding Doppler images. The location of
the main components in the system, such as the Roche lobe of the
secondary star and the predicted gas stream trajectory and the
Keplerian trajectory of the disk along the stream are indicated. These
tracks were calculated adopting the values $K_2 \sim 437$~\kms and
$q=M_2/M_1 \sim 0.06$. 

Although the disk activity seems to be lower than in previous studies
(see Section~\ref{secew}), we can see a clear detection of the bright
spot between the velocity trajectories of the gas stream and of the
disk along the stream, as previously reported in \citet{mrw94,nsv08}
and in other systems. 

We must note that the Doppler maps shown in Fig.~\ref{figmap} were
constructed without subtracting any K-type template spectrum from the
original data. We note here the presence of intense H$\alpha$ emission
exactly at the position of the secondary star. 
The emission coming from the secondary star can be also identified
through the short tracks of the expected S-wave emission on the
trailed spectrogram.  

The H$\beta$ Doppler map also shows emission around the position of
the Roche lobe but it is less clear. The S/N of the spectra at
H$\beta$ is significantly lower and the line is also close the edge of
the CCD which makes this map somewhat uncertain.

\section{Narrow H$\alpha$ component emission from the secondary
star\label{secha}}

\citet{cas97} already found a narrow H$\alpha$ emission line
associated to the donor star in the spectra of the black hole binary
Nova Muscae 1991. This line was tentatively associated to
chromospheric activity of the secondary star in this 
system. It is known that rapidly rotating stars as young T-Tauri stars
\citep[e.g.][]{aam89} and/or old, non-interacting RS CVn binaries
\citep[e.g.][]{her85,fmh86} show a high level of chromospheric
activity, with significant filling-in of the H$\alpha$ core and,
occasionally, reversal into emission. Rapidly rotating and convective
stars are capable of manifesting surface magnetic fields through the
dynamo process \citep{par55}. The induced magnetic fields lead to
confinement and heating of plasma, which produce chromospheric and
coronal emissions (e.g \ion{Ca}{ii} H and K lines, H$\alpha$ and
X-rays). Observations of these phenomena in solar type stars have shown
that the chromospheric activity increases with rotation rate
\citep[e.g.][]{noy84}.

Companion stars of X-ray transients are tidally locked, with short
orbital periods and, therefore, large rotational velocities. Thus, it
is not surprinsing that the secondary star in \axb, with $v\sin
i\sim83$~\kmso, is chromospherically active.  However, this is the
first time that chromospheric emission from the secondary star has
been clearly detected in this system before subtracting the spectrum of a
template K-type star. Note that \citet{mrw94} do detect a narrow
S-wave in their trailed spectra but only after subtracting the
template star, which, as a matter of fact, provides a limit on the
chromospheric emission set by the equivalent width of the H$\alpha$
absorption in the template star. 
The reason this chromospheric component has not been detected in
previous studies seems to hold on variability arguments, perhaps 
related to the low level of accretion disc activity seen in our 
data (which ortherwise might somehow dilute the chromospheric
component) or simply due to variations in the donor star's activity
(magnetic cycles).

\begin{figure*}[ht!]
\centering
\includegraphics[width=12.5cm,angle=90]{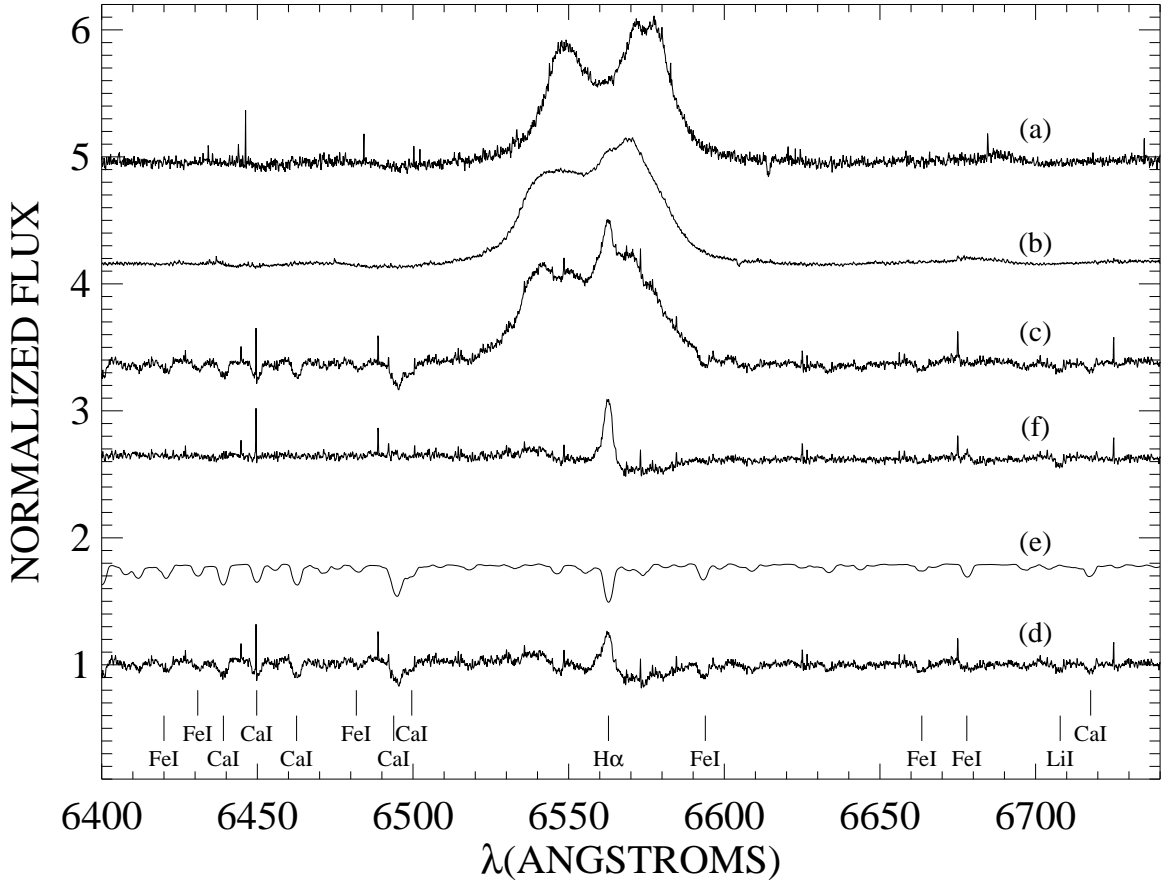}
\caption{\footnotesize{Isolating the H$\alpha$ narrow component in
emission coming from the secondary star in \axb\ from the H$\alpha$
profiles of the accretion disk. The spectra displayed are from top to
bottom: (a) Average H$\alpha$ profile in the rest frame of the
system center-of-mass; (b) Doppler-corrected average of the spectrum
(1) in the rest frame of the secondary star; (c) Doppler corrected
average spectrum of the secondary star;{\rm (d)$=$(c)-(b) } spectrum of
the secondary star without emission lines from the accretion disk,
showing the narrow emission H$\alpha$ component of the secondary star;
(e) spectrum of the K-type template star properly broadened; and 
{\rm (f)$=$(d)-0.76(e) }is chromospheric spectrum of the secondary star.
{\rm Arbitrary offsets have been applied to the spectra for the sake
of clarity. }
}}     
\label{figha}
\end{figure*}

We attempted to detrend the narrow H$\alpha$ component from the broad
underlying accretion disk contribution. The K3-4.5V template spectrum
used in this work shows an H$\alpha$ line in absorption of
$\sim1.3-1.4$~{\AA}.
This could be considered as a lower limit to
the EW of the chromospheric emission of the secondary star in \axb. 
In fact, the derived effective temperature of \teff$\sim 4900$~K by
\citet{gon04} favours a slightly earlier spectral type, i.e. K2V,
which would shift this limit up to 1.4--1.5~{\AA}.

In the Doppler corrected average of the 20 spectra of the system, this
narrow component appears as a narrow feature on top of the broad
doubled-peak H$\alpha$ emission produced by the accretion disk (see
spectrum (c) of Fig.~\ref{figha}). The emission of the accretion disk
can be approximated by a Doppler-shifted average (in the rest frame of
system center-of-mass) of the double-peaked profile (spectrum (a) of 
Fig.~\ref{figha}). This is not exactly the real accretion disk
emission profile since it also contains the smeared S-wave component
of the secondary star. We then shifted in velocity 20 versions of
spectrum (a) to the rest frame of the secondary star at the given
phase of each individual spectrum and computed again the average using
the same weights as before. The resulting profile (spectrum (b) of
Fig.~\ref{figha}) was subsequently subtracted from the original 
Doppler corrected average, i.e. spectrum (c), leaving completely 
isolated the narrow H$\alpha$ component associated with the secondary
star (spectrum (d) of Fig.~\ref{figha}). After this subtraction
we added a constant value equal to 1 since the continuum level of the
resulting spectrum was at zero. We did this to avoid to work with
negative spectral points. We performed a second
iteration of the whole process by firstly subtracting each
original spectrum by spectrum (d), properly corrected by the star's
velocity at each orbital phase, in order to eliminate the smeared
narrow component of the secondary star from the profiles 
(a) and (b). In fact, the spectra displayed in
Fig.~\ref{figha} shows the results of this last iteration. 

The spectrum (d) of Fig.~\ref{figha} is the normalized spectrum of 
the secondary star without the accretion disk contribution.
We measure the equivalent width of the narrow H$\alpha$ component 
from the secondary star in this spectrum to be 
$1.40 \pm 0.29$ {\AA}. The error bar has been estimated by 
changing the position of the continuum, taking into account the S/N
ratio of the spectrum of the secondary star. 
In order to measure the real chromospheric emission H$\alpha$ line,
one needs to subtract the photospheric component using a template star
with the same spectral type \citep[see e.g.][]{mon95}.
We then subtracted a K-type template star properly broadened (spectrum
(e)) and scaled by a veiling factor of 0.76, and find an EW of
$2.15 \pm 0.26$ {\AA} (spectrum (f)), which has to be corrected for 
the same veiling factor, leading to a final value of EW(H$\alpha) =
2.82 \pm 0.34$ {\AA}.  
 
\begin{table*}
\caption[]{H$\alpha$ fluxes and equivalent widths, and Rossby numbers 
of the LMXBs displayed in Fig.~\ref{figrharo}.}    
\label{tabfluxha}
\centering
\begin{tabular}{lrrrrrrrrr}
\hline
\hline
\noalign{\smallskip}
Object & EW(H$\alpha$)$^{a}$ & \teff & \logg & $\log F({\rm H}\alpha)$ & $\log R({\rm H}\alpha)$ & $\tau_{c}$$^{b}$ & $P_{\rm orb}$ & $\log R_0$ & Refs.$^{c}$\\       
\noalign{\smallskip}
   & [\AA] & [K] & [dex] & [$10^6$ erg cm$^{-2}$ s$^{-1}$] & [dex] & [days] & [days] & [dex] & \\       
\noalign{\smallskip}
\hline
\noalign{\smallskip}
A0620--00         & $2.82\pm0.34$ & $4900\pm100$ & 4.2 & $6.95\pm0.05$ & $-3.57\pm0.06$ & 22.7 & 0.32 & -1.85 & 1,2 \\
Nova Muscae 1991  & $3.65\pm0.32$ & $4500\pm100$$^{d}$ & 4.1$^{d}$ & $6.89\pm0.08$ & $-3.48\pm0.09$ & 24.6 & 0.43 & -1.74 & 3,4 \\
Cen X-4$^{e}$     & $2.27\pm0.67$ & $4500\pm100$ & 3.9 & $6.68\pm0.11$ & $-3.68\pm0.12$ & 24.7 & 0.63 & -1.58 & 5,6 \\
Cen X-4$^{f}$     & $4.40\pm0.50$ & $4500\pm100$ & 3.9 & $6.96\pm0.05$ & $-3.40\pm0.06$ & 24.7 & 0.63 & -1.58 & 6,7 \\
Nova Scorpii 1994 & $10.0\pm0.50$ & $6100\pm100$ & 3.7 & $7.92\pm0.02$ & $-2.97\pm0.04$ & 7.6  & 2.62 & -0.60 & 8,9,10 \\
\noalign{\smallskip}
\hline     
\end{tabular}
\begin{list}{}{}
\item[$^{a}$] Corrected equivalent width of narrow H$\alpha$ component
of the secondary star.
\item[$^{b}$] Convective turnover time derived from the equations in 
\citet{noy84} for dwarfs stars and using the turnover times given in
\citet{bas87} for subgiant and giant stars.
\item[$^{c}$] References: (1) \citet{gon04}, (2) \citet{mar86}, 
(3) \citet{cas97}, (4) \citet{rem92}, (5) \citet{tor02},
(6) \citet{gon05}, (7) \citet{dav05}, (8) \citet{sha99}, 
(9) \citet{gon08a}, (10) \citet{hoo98}.
\item[$^{d}$] Adopted values according to the effective temperature and surface 
gravity for the spectral type and orbital period.
\item[$^{e}$] H$\alpha$ equivalent width obtained from \citet{tor02}
\item[$^{f}$] H$\alpha$ equivalent width obtained from \citet{dav05} 
\end{list}
\end{table*}

\section{Discussion}

\citet{str90} measured the H$\alpha$
EWs of a sample of F6-M2 single and binary stars, showing that the
H$\alpha$ EW increases towards shorter rotation periods, $P$.
Extrapolation of this trend EW vs. $P$ at the orbital period of \axb\
provides an expected EW of $\sim 2.3$~{\AA}, assuming that the
secondary star is a dwarf main-sequence star and its rotation is
sychronized with the orbital motion. This EW is
marginally consistent (at 1.5$\sigma$) with the observed EW of 
the secondary star.
However, the extrapolation might
not be adequate since chromospheric features powered by rotation
typically saturates for period shorter than 1--3 days
\citep[e.g.][for chromospheric \ion{Mg}{ii} lines]{car07}. 

\begin{figure}[ht!]
\centering
\includegraphics[width=8.5cm,angle=0]{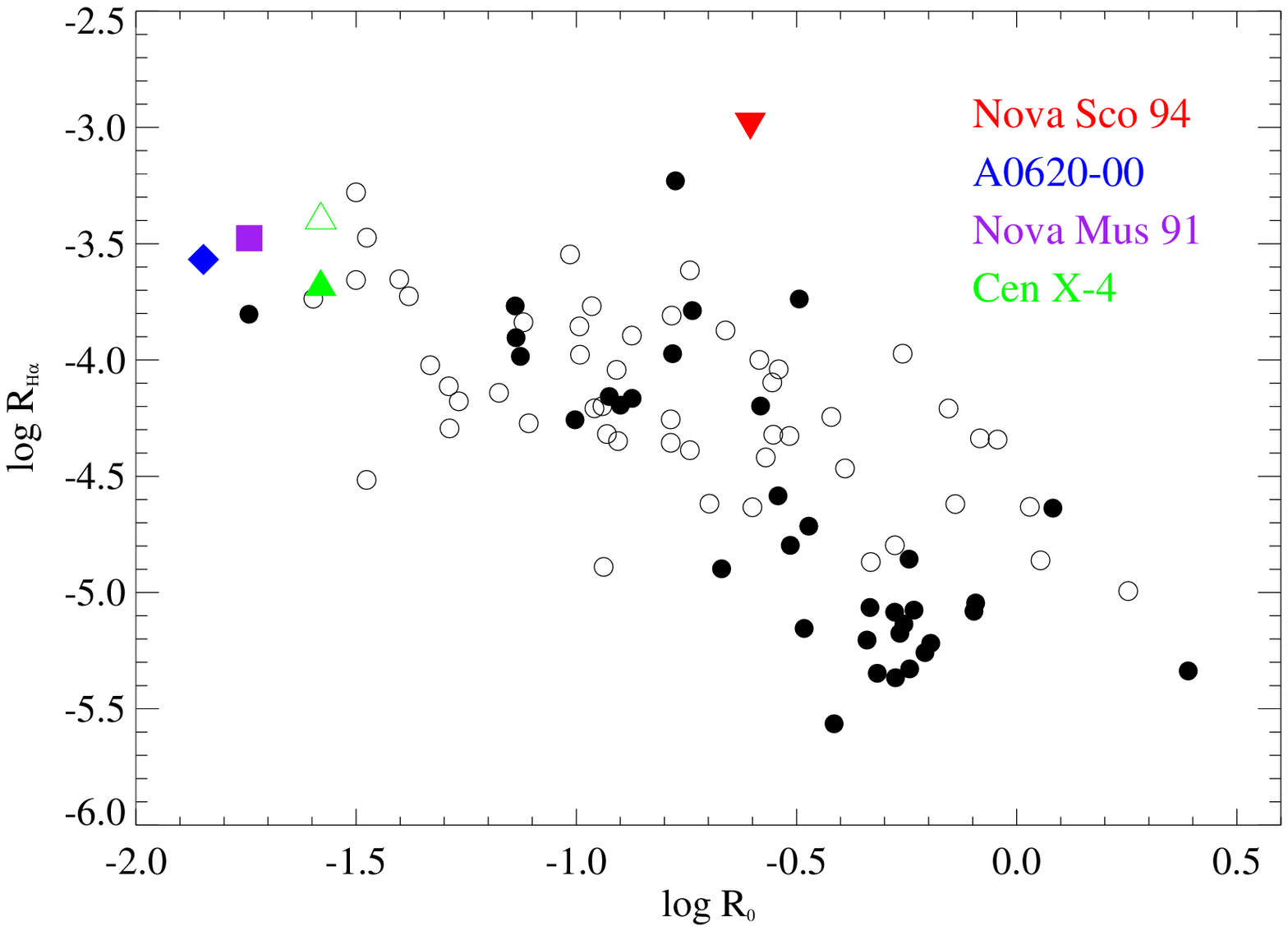}
\caption{\footnotesize{Chromospheric flux ratios vs Rossby numbers of
chromospherically active single stars from \citet[][in press, filled
circles]{lop10} and chromospherically active binary
systems (RS CVn and BY Dra classes) from \citet[][open
circles]{mon95}. Low mass X-ray binaries are also depicted: A0620--00
(diamond), Nova Muscae 1991 (square), Cen X-4 (filled triangle, from
Torres et al. 2002; open triangle, from D'avanzo et al. 2005), and
Nova Scorpii 1994 (inverted triangle).}}      
\label{figrharo}
\end{figure}

We can also compare the H$\alpha$ EW, converted into flux following
the approach of \citet{sod93}, with the observations of field stars
and binaries. Thus, the H$\alpha$ flux at the stellar surface,
is $F_{\rm H\alpha}=8.8\pm1.1\times10^6$ erg~cm$^{-2}$~s$^{-1}$ 
(see Table~\ref{tabfluxha}). We use a similar prescription than that
of Equation~2 in \citet{sod93}, i.e. $F_{\rm H\alpha} = {\rm EW
(H\alpha)} F_c$, where is $F_c$ is the continuum flux at H$\alpha$
 and is derived using the flux calibration of \citet{hal96}, 
 $\log F_c = 7.538 - 1.081 (B-V)_0$.
Here we have used $(B-V)_0=0.965$ estimated from theoretical colours
\citep{bes98} according to the stellar parameters of the secondary
star \citep{gon04}. The surface flux is usually normalized to the
bolometric flux, i.e. $R_{\rm H\alpha}=F_{\rm H\alpha}/\sigma T_{\rm
eff}^4$. The connection between chromospheric activity and rotation is
obtained by studying the correlation $R_{\rm H\alpha}$ with the {\em
Rossby} number $R_0=P/\tau_c\lesssim 2\pi R_\star/\tau_c v \sin i$, 
where $\tau_c$ is the convective turnover time, $P$, the rotation 
period of the secondary star and $R_\star$, its radius. Using the 
{\em Rossby} number is usually preferred over $P$ and $v \sin i$ 
since it does not depend on the mass of the star.

For the case of \axb, we adopt $T_{\rm eff}=4900$~K \citep{gon04},  
$P=P_{\rm orb}=0.32$~d and $\tau_c\sim23$~d, computed from Eq.~(4) of
\citet{noy84}. We derive $\log~R_{\rm H\alpha}=-3.57\pm0.06$ and
$\log~R_0=-1.85$, and this is listed in Table~\ref{tabfluxha},
together with values derived for other quiescent X-ray transients. 
In Fig.~\ref{figrharo} we compare these values with those of
chromospherically active single stars \citep[][in press]{lop10}
and binary systems (RS CVn and BY Dra classes) from \citet[][]{mon95}.
Our value of the {\em Rossby} number places the secondary star of 
\axb\ in the region of activity saturation, where all measurements
tend to the same average value of $\log~R_{\rm H\alpha}\sim-3.7$, and
is consistent with the general trend. 
The other X-ray binaries show similar results, except for the black
hole X-ray binary Nova Scorpii 1994 which displays a too large $R_{\rm
H\alpha}$ value for its relatively low Rossby number. This is 
even more evident when comparing the H$\alpha$ fluxes, with $\log
F_{\rm H\alpha}\sim7.9$ for Nova Scorpii 1994, significantly higher,
by almost one order of magnitude, than the saturation level
at $\log F_{\rm H\alpha}\sim6.9$. In addition, \citet{dav05} also
suggested that the H$\alpha$ EW of the secondary star in
Cen X-4 is correlated with the veiling of the accretion disk (see
Table~\ref{tabfluxha}), by comparing their values with those given by
\citet{tor02}.  
Although this is not expected in a chromospheric activiy scenario, the
two Cen X-4 points in Fig.~\ref{figrharo} fall in the saturation
region, together with other X-ray and chromospherically active
binaries. The behaviour seen in Nova Scorpii 1994 and Cen X-4 
suggests that rapid rotation might not be the only explanation for
the narrow H$\alpha$ feature, at least in these LMXBs, but 
perhaps a combination of rotation and reprocessing of X-ray flux from
the accretion disk into H$\alpha$ photons in the secondary star.
Hence, it is worth investigating in \axb\ if X-ray heating could 
be an alternative explanation for this feature. 

The system \axb\ has been observed in quiescence with the {\em
Chandra} X-ray satellite, providing a 0.5-10~KeV unabsorbed flux
$F_{X,0}=6.7^{+0.8}_{-2.3}\times10^{-14}$~erg~cm$^{-2}$~s$^{-1}$
\citep{gal06}. 

The X-ray flux at the stellar surface can be computed as
$F_{X,\star}=F_{X,0}(d/a)^2=9.5\times10^6$~erg~cm$^{-2}$~s$^{-1}$, 
where we have adopted an orbital separation $a=4.47$~\Rsun and a
distance $d=1.2$~kpc. This means that almost 92\% of the incident
X-ray radiation would have to be reprocessed to  
H$\alpha$ photons in order to power the observed H$\alpha$ emission.
Following \citet{hyn02}, $F_{\rm H\alpha,\star}=f_1f_2F_{X,\star}$,
where $f_1$ is the fraction of X-ray emission intercepted by the
companion, i.e. the solid angle subtended by the companion 
from the compact object ($f_1=[R_\star/(2a)]^2$), and $f_2\lesssim
0.3$ is the fraction of input energy emitted in H$\alpha$ 
\citep[][and references therein]{hyn02}. Adopting
$R_\star=1.1$~\Rsun from \citet{gon04}, we obtain $F_{\rm
H\alpha,\star} \lesssim 4.5 \times 10^{-3} F_{X,\star}$. This number 
is significantly lower than the observed value, what indicates that 
the incident X-ray irradiation is not enough to produce the narrow 
H$\alpha$ line in the secondary star. \citet{dav05} also derived these 
quantities for the case of the neutron star binary Cen X-4 and found 
both estimates to be consistent, due to the fact that the X-ray flux 
in Cen X-4 is $F_{X,\rm Cen X-4}=5\times 10^8$~erg~cm$^{-2}$~s$^{-1}$, 
i.e. almost two order of magnitude higher than in \axb. 

There is still a remote posibility that the source of irradiating  
photons is hidden away in the EUV (Extreme UV) energy range, between  
100--1200~{\AA}. Although this energy range is not directly observed, we  
can roughly guess how much flux is involved through interpolating the  
nearby soft X-ray and Far-UV (FUV) emission.
The Far-UV flux (in the range 1350--2200~{\AA}) has been determined at 
$F_{{\rm FUV},0}=0.2-1.4\times10^{-13}$~erg~cm$^{-2}$~s$^{-1}$ 
\citep{mhr95}, i.e. similar to the X-ray flux.
This together with the absence of \ion{He}{ii}~$\lambda$4686~{\AA}
line emission in the optical spectrum \citep{mrw94} suggests that 
flux in the EUV should be of the order of
$10^{-13}$~erg~cm$^{-2}$~s$^{-1}$. Even 
if we consider all the ionizing photons (X-ray$+$EUV$+$FUV), 
the total flux would be roughly three times the X-ray flux, i.e. 
$\sim 2\times10^{-13}$~erg~cm$^{-2}$~s$^{-1}$. 
Therefore, if we assume that the incident irradiation is
$F_{X-UV,\star}= 3\,F_{X,\star}$, then the 31\% of the incident 
radiation would have to be reprocessed to  
H$\alpha$ photons in order to power the observed H$\alpha$ emission,
which still is a too large fraction compared to the
fraction previously estimated to be bellow 1\%.

\section{Summary}

We have presented high-resolution UVES/VLT spectroscopy 
of the black hole binary \axb\ at quiescence. 
Our orbital parameters are consistent with previous studies by 
\citet{mrw94} and \citet{nsv08}. In particular, we derive $P_{\rm
orb}=0.32301405(1)$ d and
$K_2=437.1\pm2.0$ \kmso. 
These values, together with the mass ratio
$q=M_2/M_1=0.062\pm0.010$, implies a minimum mass for the
compact object of $M_1 \sin^3 i = 3.15 \pm 0.10$~\Msuno. 

We also performed a Doppler tomography of the accretion disk emission
and we discover emission at the position of the
secondary star in the Doppler maps of H$\alpha$ and H$\beta$, not
detected in previous studies. We isolate the chromospheric H$\alpha$
emission from the secondary star and measure an equivalent width of
$2.82\pm0.34$~{\AA}. This equivalent width is too large to be
explained by X-ray and/or UV irradiation from the inner 
accretion disk and, therefore, chromospheric activity, induced 
by rapid rotation, seems the most likely origin of this feature in 
the black hole binary \axb. 

\begin{acknowledgements}

J.I.G.H. acknowledges support from the project AYA2008-00695 of the
Spanish Ministry of Education and Science. J.C. acknowledges support 
from the Spanish Ministry of Science and Technology through the project
AYA2007-66887. This work has been partially funded by the Spanish
MICINN under the Consolider-Ingenio 2010 Program grant CSD2006-00070:
First Science with the GTC (http://www.iac.es/consolider-ingenio-gtc).  
We are grateful to Tom Marsh for the use
of the MOLLY analysis package. J.I.G.H. is grateful to Javier L\'opez
Santiago for providing us with the data of chromospherically active
single and binary stars. J.I.G.H. also thanks helpful discussions with
Javier L\'opez Santiago, David Montes and Raquel Mart\'inez Arn\'aiz. 

\end{acknowledgements}


\begin{thebibliography}{}

\bibitem[Al-Naimiy(1978)]{al78} 
Al-Naimiy, H.M. \ 1978, \apss, 53, 181 

\bibitem[Appenzeller \& Mundt(1989)]{aam89} 
Appenzeller, I., \& Mundt, R.\ 1989, \aapr, 1, 291 

\bibitem[Basri(1987)]{bas87} 
Basri, G.\ 1987, \apj, 316, 377 

\bibitem[Bessell et al.(1998)]{bes98} 
Bessell, M.~S., Castelli, F., \& Plez, B.\ 1998, \aap, 333, 231 

\bibitem[Cardini \& Cassatella(2007)]{car07} 
Cardini, D., \& Cassatella, A.\ 2007, \apj, 666, 393 

\bibitem[Cantrell et al.(2008)]{can08} 
Cantrell, A.~G., Bailyn, C.~D., McClintock, J.~E., \& Orosz, J.~A.\
2008, \apjl, 673, L159  

\bibitem[Cantrell et al.(2010)]{can10} 
Cantrell, A.~G. et al.\ 2010, \apj, submitted (arXiv:1001.0261)

\bibitem[Casares et al.(1996)]{cas96}
Casares, J., Mouchet., M.,  Mart\'\i{}nez-Pais, I.G., \& Harlaftis, E.T.\
1996, MNRAS, 282, 182  

\bibitem[Casares et al.(1997)]{cas97}
Casares, J., Martin, E.~L., Charles, P.~A., Molaro, P., \& Rebolo, R.\
1997, New Astronomy, 1, 299  

\bibitem[Casares et al.(2007)]{cas07}
Casares, J., Bonifacio, P., Gonz{\'a}lez Hern{\'a}ndez, J.~I., Molaro,
P., \& Zoccali, M.\ 2007, \aap, 470, 1033  

\bibitem[Casares et al.(2009)]{cas09} 
Casares, J., Gonz{\'a}lez Hern{\'a}ndez, J.~I., Israelian, G., \&
Rebolo, R.\ 2009, \mnras, in press

\bibitem[Collins \& Truax(1995)]{col95}
Collins, G.W.II \& Truax, R.J.\ 1995, ApJ, 439, 860

\bibitem[D'Avanzo et al.(2005)]{dav05} 
D'Avanzo, P., Campana, S., Casares, J., Israel, G.~L., Covino, S., 
Charles, P.~A., \& Stella, L.\ 2005, \aap, 444, 905 

\bibitem[Elvis et al.(1975)]{elv75} 
Elvis, M., Page, C.~G., Pounds, K.~A., Ricketts, M.~J., \& Turner,
M.~J.~L.\ 1975, \nat, 257, 656  

\bibitem[Fekel et al.(1986)]{fmh86} 
Fekel, F.~C., Moffett, T.~J., \& Henry, G.~W.\ 1986, \apjs, 60, 551 

\bibitem[Froning et al.(2007)]{fro07} 
Froning, C.~S., Robinson, E.~L., \& Bitner, M.~A.\ 2007, \apj, 663,
1215  

\bibitem[Gallo et al.(2006)]{gal06} 
Gallo, E., Fender, R.~P., Miller-Jones, J.~C.~A., Merloni, A., Jonker,
P.~G., Heinz, S., Maccarone, T.~J., \& van der Klis, M.\ 2006, \mnras,
370, 1351  

\bibitem[Gelino et al.(2001)]{gel01} 
Gelino, D.~M., Harrison, T.~E., \& Orosz, J.~A.\ 2001, \aj, 122, 2668

\bibitem[Gelino et al.(2006)]{gel06}
Gelino, D. M., Balman, \c{S}., Kililo\u{g}lu, \"U., Yilmaz, A.,
Kalemci, E., \& Tomsick, J. A. 2006, \apj, 642, 438

\bibitem[Gonz{\'a}lez Hern{\'a}ndez et al.(2004)]{gon04} 
Gonz{\'a}lez Hern{\'a}ndez, J.~I., Rebolo, R., Israelian, G.,
Casares, J., Maeder, A., \& Meynet, G.\ 2004, \apj, 609, 988 

\bibitem[Gonz{\'a}lez Hern{\'a}ndez et al.(2005)]{gon05} 
Gonz{\'a}lez Hern{\'a}ndez, J.~I., Rebolo, R., Israelian, G., 
Casares, J., Maeda, K., Bonifacio, P., \& Molaro, P.\ 2005, 
\apj, 630, 495 

\bibitem[Gonz{\'a}lez Hern{\'a}ndez et al.(2006)]{gon06} 
Gonz{\'a}lez Hern{\'a}ndez, J.~I., Rebolo, R., Israelian, G.,
Harlaftis, E.~T., Filippenko, A.~V., \& Chornock, R.\ 2006, \apjl,
644, L49  

\bibitem[Gonz{\'a}lez Hern{\'a}ndez et al.(2008a)]{gon08a} 
Gonz{\'a}lez Hern{\'a}ndez, J.~I., Rebolo, R., \& Israelian, G.\
2008a, \aap, 478, 203  

\bibitem[Gonz{\'a}lez Hern{\'a}ndez et al.(2008b)]{gon08b}
Gonz{\'a}lez Hern{\'a}ndez, J.~I., Rebolo, R., Israelian, G.,
Filippenko, A.~V., Chornock, R., Tominaga, N., Umeda, H., \& Nomoto, 
K.\ 2008b, \apj, 679, 732  

\bibitem[Hall(1996)]{hal96} 
Hall, J.~C.\ 1996, \pasp, 108, 313 

\bibitem[Harrison et al.(2007)]{har07} 
Harrison, T.~E., Howell, S.~B., Szkody, P., \& Cordova, F.~A.\ 2007,
\aj, 133, 162  

\bibitem[Harlaftis et al.(1997)]{har97} 
Harlaftis, E.~T., Steeghs, D., Horne, K., \& Filippenko, A.~V.\ 1997,
\aj, 114, 1170 

\bibitem[Harlaftis et al.(1999)]{har99} 
Harlaftis, E., Collier, S., Horne, K., \& Filippenko, A.~V.\ 1999, 
\aap, 341, 491 

\bibitem[Herbig(1985)]{her85} 
Herbig, G.~H.\ 1985, \apj, 289, 269 

\bibitem[van der Hooft et al.(1998)]{hoo98} 
van der Hooft, F., Heemskerk, M.~H.~M., Alberts, F., \& van Paradijs, 
J.\ 1998, \aap, 329, 538 

\bibitem[Hynes et al.(2002)]{hyn02} 
Hynes, R.~I., Zurita, C., Haswell, C.~A., Casares, J., Charles, 
P.~A., Pavlenko, E.~P., Shugarov, S.~Y., \& Lott, D.~A.\ 2002, 
\mnras, 330, 1009
 
\bibitem[Hynes et al.(2005)]{hyn05} 
Hynes, R.~I., Robinson, E.~L., \& Bitner, M.\ 2005, \apj, 630, 405 

\bibitem[L\'opez-Santiago et al.(2010)]{lop10} 
L\'opez-Santiago, J. et al.\ 2010, \aap, in press

\bibitem[Marsh \& Horne(1988)]{mah88} 
Marsh, T.~R., \& Horne, K.\ 1988, \mnras, 235, 269 

\bibitem[Marsh et al.(1994)]{mrw94} 
Marsh, T.~R., Robinson, E.~L., \& Wood, J.~H.\ 1994, \mnras, 266, 137

\bibitem[McClintock \& Remillard(1986)]{mar86} 
McClintock, J.~E., \& Remillard, R.~A.\ 1986, \apj, 308, 110 

\bibitem[McClintock et al.(1995)]{mhr95} 
McClintock, J.~E., Horne, K., \& Remillard, R.~A.\ 1995, \apj, 442,
358  

\bibitem[Montes et al.(1995)]{mon95} 
Montes, D., Fernandez-Figueroa, M.~J., de Castro, E., \& Cornide, M.\
1995, \aaps, 109, 135  

\bibitem[Murdin et al.(1980)]{mur80} 
Murdin, P., Allen, D.~A., Morton, D.~C., Whelan, J.~A.~J., \& Thomas,
R.~M.\ 1980, \mnras, 192, 709  

\bibitem[Neilsen et al.(2008)]{nsv08} 
Neilsen, J., Steeghs, D., \& Vrtilek, S.~D.\ 2008, \mnras, 384, 849 

\bibitem[Noyes et al.(1984)]{noy84}
Noyes, R.~W., Hartmann, L.~W., Baliunas, S.~L., Duncan, D.~K., 
\& Vaughan, A.~H.\ 1984, \apj, 279, 763 

\bibitem[Oke(1977)]{oke77} 
Oke, J.~B.\ 1977, \apj, 217, 181 

\bibitem[Orosz et al.(1994)]{oro94} 
Orosz, J.~A., Bailyn, C.~D., Remillard, R.~A., McClintock, J.~E., 
\& Foltz, C.~B.\ 1994, \apj, 436, 848 

\bibitem[Parker(1955)]{par55} 
Parker, E.~N.\ 1955, \apj, 122, 293

\bibitem[Remillard et al.(1992)]{rem92} 
Remillard, R.~A., McClintock, J.~E., \& Bailyn, C.~D.\ 1992, 
\apjl, 399, L145 

\bibitem[Shahbaz et al.(1999)]{sha99} 
Shahbaz, T., van der Hooft, F., Casares, J., Charles, P.~A., 
\& van Paradijs, J.\ 1999, \mnras, 306, 89 

\bibitem[Shahbaz et al.(1994)]{sha94} 
Shahbaz, T., Naylor, T. \& Charles, P.~A.\ 1994, \mnras, 268, 756 

\bibitem[Shahbaz et al.(2004)]{sha04} 
Shahbaz, T., Hynes, R.~I., Charles, P.~A., Zurita, C., Casares, J.,
Haswell, C.~A., Araujo-Betancor, S., \& Powell, C.\ 2004, \mnras, 354,
31  

\bibitem[Soderblom et al.(1993)]{sod93} 
Soderblom, D.~R., Stauffer, J.~R., Hudon, J.~D., \& Jones, B.~F.\
1993, \apjs, 85, 315  

\bibitem[Strassmeier et al.(1990)]{str90}
Strassmeier, K.~G., Fekel, F.~C., Bopp, B.~W., Dempsey, R.~C., 
\& Henry, G.~W.\ 1990, \apjs, 72, 191 

\bibitem[Torres et al.(2002)]{tor02} 
Torres, M.~A.~P., Casares, J., Mart{\'{\i}}nez-Pais, I.~G., 
\& Charles, P.~A.\ 2002, \mnras, 334, 233 

\bibitem[Wade \& Horne(1988)]{wah88} 
Wade, R.~A., \& Horne, K.\ 1988, \apj, 324, 411 

\end{thebibliography}
\end{document}